\documentclass[12pt, preprint]{aastex}

\newcommand{\flux}{ergs s$^{-1}$ cm$^{-2}$}
\newcommand{\lum}{ergs s$^{-1}$}
\newcommand{\lfive}{\ensuremath{L_{5100}}}
\def\ucre{${\rm cts~s^{-1}}$}

\newcommand{\etal}{et al.}

\newcommand{\msun}{\ensuremath{M_{\odot}}}

\newcommand{\kms}{km s\ensuremath{^{-1}}}

\newcommand{\si}[1]{\ensuremath{_{\textrm{\scriptsize{#1}}}}}

\slugcomment{To Appear in ApJ (2007)} \shorttitle{Dwarf Disk Galaxy
with an IMBH} \shortauthors{DONG ET AL.}

\begin{document}

\title{SDSS J160531.84+174826.1: A Dwarf Disk Galaxy With An Intermediate-Mass
Black Hole}

\author{Xiaobo Dong\altaffilmark{1,2},Tinggui Wang\altaffilmark{1,2},
Weimin Yuan\altaffilmark{3}, Hongguang Shan\altaffilmark{3}, Hongyan
Zhou\altaffilmark{1,2,4}, Lulu Fan\altaffilmark{1,2}, Liming
Dou\altaffilmark{3}, Huiyuan Wang\altaffilmark{1,2}, Junxian
Wang\altaffilmark{1,2}, and Honglin Lu\altaffilmark{1,2} }

\email{xbdong,twang@ustc.edu.cn; wmy@ynao.ac.cn}
\altaffiltext{1}{Center for Astrophysics, University of Science and
Technology of China, Hefei, Anhui, 230026, China}
\altaffiltext{2}{Joint Institute of Galaxies and Cosmology, Shanghai
Observatory and University of Science and Technology of China}
\altaffiltext{3}{National Astronomical Observatories/Yunnan
Observatory, Chinese Academy of Sciences, P.O. Box 110, Kunming,
Yunnan 650011, China} \altaffiltext{4}{Department of
Astronomy, University of Florida, Gainesville, FL 32611}

\begin{abstract}
We report the discovery of a dwarf Seyfert 1 active galactic nucleus
(AGN) with a candidate intermediate-mass black hole hosted by the
dwarf galaxy SDSS J160531.84+174826.1 at $z=0.032$. A broad
component of the H$\alpha$ line with FWHM=781\kms~is detected in its
optical spectrum, and a bright, point-like nucleus is evident from a
HST imaging observation. Non-thermal X-ray emission is also detected
from the nucleus. The black hole mass, as estimated from the
luminosity and width of the broad H$\alpha$ component, is about
$7\times 10^4 \msun$. The host galaxy appears to be a disk galaxy
with a boxy bulge or nuclear bar; with an absolute magnitude of $M_R
= -17.8$, it is among the least luminous host galaxies ever
identified for a Seyfert 1.

\end{abstract}

\keywords{galaxies: active --- galaxies: dwarf --- galaxies:
individual (SDSS J160531.84+174826.1) --- galaxies: nuclei ---
galaxies: Seyfert}

\section{Introduction}

Supermassive black holes (SMBHs) with masses $M_{BH} \gtrsim 10^6
\msun $ have been convincingly inferred to be present in the centers
of nearby galaxies (see Kormendy 2004 for a review), and are
generally believed to reside in most, if not all, galaxies with a
spheroidal stellar component (Kormendy \& Gebhardt 2001; Ferrarese
\& Ford 2005). However, little is known about their low mass
counterparts, i.e. black holes with masses in the range of
$10^3-10^6\msun$, presumably in the centers of (dwarf) galaxies.
These intermediate-mass black holes (IMBHs) may provide the `missing
link' in understanding the formation and evolution of SMBH seen
today. In current models of galaxy evolution in a hierarchical
cosmology, SMBH must have formed from much less massive ``seed''
black holes and grown up by accretion and/or merging. It is likely
that there exists a population of IMBHs in the present universe,
that have not the opportunity to be full-grown (e.g., Islam, Taylor
\& Silk 2004).

The search for IMBH turns out to be a difficult task, however, since
IMBHs are beyond the reach of direct measurement using star/gas
dynamics, by which most nearby SMBH are unveiled (see van der Marel
2004 for a review). The most promising approach is to search for
dwarf active galactic nuclei (AGN) that are hosted in small
galaxies---a scaled-down version of typical Seyfert galaxies and
quasars. So far, convincing evidence for the existence of IMBHs has
been found in only two AGN, NGC~4395 (Filippenko \& Ho 2003,
hereafter FH03; $M_{BH} = (3.6\pm1.1) \times 10^5 \msun$, Peterson
et al. 2005, hereafter P05) and Pox 52 ($M_{BH} \approx 1.6 \times
10^5 \msun$, Barth et al. 2004, hereafter B04), with NGC~4395 being
the only one having accurate black hole mass measurement via
reverberation mapping. In addition, Greene \& Ho (2004; hereafter
GH04) found 19 IMBH candidates from the Sloan Digital Sky Survey
(SDSS; Stoughton et al. 2002) First Data Release (Abazajian et al.
2003), whose black hole masses were estimated by using the
linewidth-luminosity-mass scaling relationship (Kaspi et al. 2000).
All these objects, except NGC 4395, have a very high accretion rate
close to the Eddington limit.

In this paper, we report the discovery of a dwarf AGN with the
estimated black hole mass as low as $\sim 10^5 \msun$, that is
hosted by SDSS J160531.84+174826.1 (hereafter J1605+1748)---a dwarf
(most likely disk) galaxy at a redshift $z=0.032$. It was found from
our on-going program of a systematic search for AGN with candidates
IMBHs or hosted in dwarf galaxies from the SDSS Fifth Data Release.
We assume a cosmology with H$_{0}$=70 km~s$^{-1}$~Mpc$^{-1}$,
$\Omega_{M} $=0.3 and $\Omega_{\Lambda}$=0.7.

\section{Data Analysis}

\subsection{Optical Spectrum}

J1605+1748 was spectroscopically observed in the SDSS on June 12,
2005 with 4048s exposure and was classified as a galaxy by the SDSS
pipeline. It has a redshift of $z=0.03167$ as determined using the
[O III]$\lambda5007$ line. Figure 1 shows the rest frame spectrum
with the Galactic reddening corrected ($E_{B-V}=0.05$, Schlegel et
al. 1998), which is dominated by starlight of the host galaxy. To
subtract the starlight and the nuclear continuum, we followed our
method\footnote{ Starlight is modeled with 6 spectral templates
which were built up from the library of simple stellar populations
of Bruzual \& Charlot (2003) by using the Ensemble Learning
Independent Component Analysis method (see Lu et al. 2006 for a
detailed description). } as described in detail in Zhou et al.
(2006). The fit was good; the standard deviation of the distribution
of the relative residuals $f_{residual}/f_{SDSS} $ in the emission
line-free region around H$\alpha$ and H$\beta$ is $\approx 0.05$,
just the noise level with respect to the signal. The left-over
emission line spectrum is plotted in Figure 1.

We fit the emission lines using the code described in detail in Dong
et al. (2005). The peaks of the [N II]$\lambda\lambda6548,6583$
doublet are well separated from the H$\alpha$ line, that shows an
apparent broad component, owing to the relatively high S/N ratio
($\gtrsim 20$) in this spectral range and the narrowness of the
lines. We model the [N II] doublet with two Gaussians with the line
ratio $\lambda$6583/$\lambda$6548 fixed to the theoretical value
2.96. We use two Gaussians to model the narrow and broad components
of H$\alpha$, assuming the narrow component has the same profile and
redshift as the [N II] doublet (see Zhou et al. 2006 for a full
account of this assumption). A good fit is achieved, with a minimum
reduced $\chi^2 = 0.73$ (d.o.f = 92), yielding a line width $794\pm
42 $\kms~FWHM for the broad H$\alpha$ component (see Figure 1). It
has been noted by V\'{e}ron-Cetty et al. (2001) that Lorentzian is a
better profile to describe the broad lines in narrow line Seyfert
1s, that have FWHMs $< 2000$\kms. In view of this argument, we use a
Lorentzian for the broad H$\alpha$ component and repeat the above
fit; this yields a similarly good fit with reduced $\chi^2 = 0.71$
(d.o.f = 92), but a model over-predicted in the [N II]$\lambda6548$
region by 7\% of the [N II]$\lambda6548$ peak height. So we adopt
the Gaussian fit. To test the reliability of this broad H$\alpha$
component, we model the whole H$\alpha$ line with a single Gaussian
with the center and width as free parameters; the result is
unacceptable with a reduced $\chi^2 = 2.75$ (d.o.f = 95) and large
residuals remained around H$\alpha$. We therefore believe that the
broad H$\alpha$ line is real. For the H$\beta$ line, since its S/N
ratio is relatively low ($\sim10$), we fix the profile parameters to
those of H$\alpha$, for both the narrow and broad components. For
the [O III]$\lambda\lambda4959,5007$ doublet, we only fit [O
III]$\lambda5007$ alone with one single Gaussian because [O
III]$\lambda4959$ is jaggy. We find the fit is good and no extended
wing for [O III]$\lambda5007$. We fit other narrow lines with a
single Gaussian profile. The results of emission line fitting are
listed in Table 1. The Balmer decrement is found to be 4.2 for the
broad lines and 4.0 for the narrow lines, indicating an extinction
color excess $E_{B-V}=$0.36 and 0.31, respectively, assuming an
intrinsic Balmer decrement of 3 (Dong et al. 2005, Zhou et al. 2006)
and a SMC-like extinction curve. The ratios of the narrow lines [O
III]$\lambda5007$/H$\beta\gtrsim$ 3 and [N II]$\lambda6583$/H$\alpha
\gtrsim 0.6$ place J1605+1748 into the AGN regime on the diagnostic
diagram (Veilleux \& Osterbrock 1987). It would be considered as a
Seyfert 1.8 in the Osterbrock (1981) classification scheme, just
like NGC 4395 (Filippenko \& Sargent 1989, hereafter FS89) and POX
52 (B04).

\subsection{Optical Image}
We retrieved an archival HST image for J1605+1748, taken with the
Wide Field Planetary Camera 2 (WFPC2) in June 1995 with a 500s
exposure. The galaxy fell on the Wide Field camera
(0\arcsec.1/pixel) as a ``bonus'' in an observation of Mkn 298
(Malkan et al. 1998). The F606W filter was used, which has a mean
wavelength of 5947\AA~and a FWHM of 1500\AA~(1997 May WFPC2 SYNPHOT
update). The data reduction was carried out following Malkan et al.
(1998). On the HST image, galaxy is well resolved thanks to the
superb spatial resolution. It is largely elongated with a major axis
of about 6\arcsec, and a bright, point-like nuclear source is
evident. We perform structural decomposition using a ``two-step''
strategy: firstly fit to the one-dimensional surface brightness
profile and then 2-D fitting to the image using the package GALFIT
(Peng et al. 2002). We construct surface brightness models, for both
the 1-D and 2-D cases, with a combination of different structural
components: a point source represented by a point-spread function
(PSF) that is generated with the Tiny Tim software (ver.
6.3)\footnote{http://www.stsci.edu/software/tinytim/} for WFPC2;
galactic component(s) parameterized by either an exponential or a
S\'{e}rsic $r^{1/n}$ function (S\'{e}rsic 1968, see Graham \& Driver
2005 for a concise review), or a combination of any of these two. We
start off the fitting procedure with the simplest one-component
model, and add in one more component only if the fit can be improved
significantly by doing so. An additional constraint comes from the
above optical spectral decomposition, which sets an upper limit on
the contribution from a point source to be $\sim15$\% the total flux
at $6000$\AA.

Firstly, the azimuthally averaged 1-D radial profile of surface
brightness is extracted using the IRAF ELLIPSE task. The best-fit
model, which is plotted in Figure 2 along with the measured profile,
is composed of a central point source plus two S\'{e}rsic
components, one dominating the inner part of the galaxy and the
other dominating the outer part. Leaving off the point source or one
of the two S\'{e}rsic gives rise to unacceptable fits with
significant and strongly structured residuals. The need for a
central point source is not surprising, as revealed from the HST
image (Figure 3). The presence of the second galactic component is
found to be significant, as shown in Figure 2. Secondly, we perform
2-D image decomposition using GALFIT, taking the results of the 1-D
fits as initial values of the fitting parameters. The best-fit model
turns out to be the same as for the 1-D fitting, but with somewhat
different parameter values, especially for the central point source.
We adopt the result from the 2-D fitting (Table 2) in this paper
because 2-D modeling can recover a nuclear source much reliable than
1-D fitting (Peng et al. 2002). Also displayed in Figure 3 are the
images of the residuals and of the model components. The fit is good
in general, with the standard deviation of the distribution for the
fractional residuals $f_{residual}/f_{HST}\sim 0.1$ within the sky
region 10 times or more brighter than the sky level. In comparison,
other models with the same or less number of model components give
rise to unacceptable fitting results that have much larger $\chi^2$
and structures in the residuals.

For easy comparisons with NGC 4395 and POX 52, We derive the Johnson
magnitudes from the F606W flux using the IRAF/SYNPHOT package. We
assume a continuum similar to that of NGC 4395 ($f_\nu \propto
\nu^{-1.5}$, Filippenko, Ho, \& Sargent 1993) for the central point
source, a template bulge spectrum (Kinney et al. 1996) for the inner
S\'ersic. We further make use of the SDSS spectrum (\S2.1) for the
outer S\'ersic since it accounts for $\gtrsim 90$ per sent of the
total F606W flux. The results are given in Table 2.

\subsection{X-ray Data}

As a remote member of the cluster Abell 2151, J1605+1748 was
observed serendipitously several times by ROSAT and XMM-Newton. We
retrieved from archives the observations with good effective
exposures, one with the ROSAT Position Sensitive Proportional
Counter (PSPC) in 1993, one with the ROSAT High Resolution Imager
(HRI) in 1997, and one with the XMM-Newton EPIC-PN\footnote{The
source position fells into a gap between CCD chips on the two
EPIC-MOS detectors. } in 2003. X-ray data reduction were performed
following the standard procedures, using the FTOOLS (ver. 6.1)
utilities and the Science Analysis System (\verb"SAS", ver 6.0.0)
for ROSAT and XMM-Newton data, respectively. Table 3 lists some
relevant parameters for the X-ray observations and data reduction.
For the XMM-Newton data, exposure periods which suffered from high
flaring background caused by soft protons are removed. Source counts
are extracted from a circle (see Table 3) and background events are
extracted from source-free regions using a concentric annulus for
PSPC and HRI, and two circles at the same CCD read-out column as the
source position for PN. In each of these observations an X-ray
source was detected at the $\gtrsim 6\sigma$ significance levels at
a position coincident with the optical point source of the galaxy
within one $\sigma$ positional error. The position of the X-ray
source detected with XMM-Newton is RA=16:05:31.87, Dec=17:48:26.11
(J2000), with an error circle of 0\arcsec.9 ($1\sigma$) in radius.

The spectral fitting is carried out using XSPEC (ver. 12.2.1). Due
to the small number of source counts, 58 for PSPC (0.1--2.4 keV) and
79 for PN (0.3-8keV), the spectra are rebinned to have a minimum of
6 counts in each energy bin, and the Cash-statistic is adopted in
the minimization instead of $\chi^2$. We fit simultaneously the two
spectra assuming the same spectral models with the parameters tied
together except the normalizations. The spectra can be fitted with a
power law with a photon index  $\Gamma =2.1 \pm 0.3$ and an
absorption column density close to the Galactic value ($3.5 \times
10^{20}$cm$^{-2}$). The unabsorbed 2--10 keV flux is $3.0\times
10^{-14}$ \flux~measured by XMM-Newton, corresponding to a
luminosity of $7.0 \times 10^{40}$ \lum. The best fit normalization
at 1 keV is a factor of 2 higher for ROSAT than for XMM-Newton. To
test the significance of this variation, we search for the maximum
confidence contours in the parameter space of the two normalizations
(two free parameters) within which they differ from each other; we
find that the variation is marginally significant at the $\sim$90\%
level. The HRI flux is found to be roughly consistent with the
XMM-Newton value assuming the same spectral shape.

\section{Results and Discussion}

In summary, observations of J1605+1748 reveal the presence of an
optical bright point source based on its HST image, a broad
H${\alpha}$ component with FWHM 781 \kms~(after correction for the
instrumental broadening of 141 \kms~FWHM) and AGN-like narrow
emission line ratios, a non-thermal X-ray luminosity ($7.0 \times
10^{40}$ \lum~in 2--10 keV) and possible X-ray variability. All
these observational facts point to the presence of a Seyfert 1 type
dwarf active nucleus in J1605+1748. This can be further supported by
the striking similarities between J1605+1748 and the two well known
dwarf AGN NGC 4395 and POX 52, whose properties are summarized in
Table 4. In fact, the AGN in J1605+1748 lies in between NGC 4395 and
POX 52 in terms of the observed luminosities of both H${\alpha}$ and
optical continuum. Furthermore, J1605+1748 is more luminous at hard
X-rays than NGC 4395 (no data available for POX 52 so far). The
equivalent width of its broad H$\beta$ component ($\sim 20$\AA,
relative to the nuclear continuum calculated from the HST F606W flux
of the point source) is typical of Seyfert 1.8/1.9 nuclei and
similar to that of NGC 4395 (FS89).

We estimate the mass of its central black hole following the widely
used linewidth-luminosity-mass scaling relation (e.g., Vestergaard
2002; McLure \& Jarvis 2002; GH04). The intrinsic monochromatic
luminosity at 5100\AA~is estimated to be $\lfive \equiv \lambda
L_{\lambda}(5100\AA)= 4.0 \times 10^{41}$ \lum~from the H$\alpha$
luminosity using the $L_{H\alpha}-\lfive$ relation given by Greene
\& Ho (2005)\footnote{This value is well consistent with that
estimated using the HST F606W flux of the point source assuming a
power-law spectrum ($f_{\lambda} \propto \lambda
^{-\alpha_{\lambda}}$), $(3.5-4.6) \times 10^{41}$ \lum~for
$\alpha_{\lambda}=0-2$.}, incorporating the extinction correction
that is derived from the Balmer decrement for the broad lines
(\S2.1). Using Equation (6) in Greene \& Ho (2005), we find a black
hole mass $6.9 \times 10^4\msun$. The uncertainty of the black hole
mass thus estimated is not well understood, however. As pointed out
by Vestergaard \& Peterson (2006), the statistical accuracy of the
masses from such a scaling relationship is a factor of $\sim 4$
($1\sigma$), and, for individual mass estimates, the uncertainty can
be as large as an order-of-magnitude. Using the
BLR-size$-$luminosity relations from Kaspi et al. (2005) and Bentz
et al. (2006), we find a black hole mass of $5.8 \times 10^4, 2.6
\times 10^5$\msun, respectively; here we adopt the formalism and
scale factor from Peterson et al. (2004) and Onken et al. (2004),
with $\lfive = 4.0 \times 10^{41}$\lum, $\sigma_{\mathrm{line}} =
332$ \kms~estimated from the broad H$\alpha$ component.

Assuming a spectral energy distribution (SED) similar to that of NGC
4395, the bolometric luminosity $L_{bol}$ for J1605+1748 is
estimated to be $\approx 3.7 \times 10^{42}$ \lum~by scaling their
\lfive~luminosities using the NGC 4395 data in P05. This is
consistent with the bolometric correction $L_{bol} \approx 9\lfive$
used for normal QSOs (Kaspi et al. 2000). Then, the Eddington ratio
$L_{bol}/L_{Edd}$ is $\approx 0.3$ for a black hole mass of $6.9
\times 10^4 \msun$. This indicates that J1605+1748 is at a high
accretion state, similar to POX 52. The broad-band spectral index
$\alpha\si{ox} \equiv
-0.3838\log[F_{\nu}(\textrm{2\,keV})/F_{\nu}(\textrm{2500\,\AA})]
\simeq -1.1$ agrees well with the extrapolation down to low
luminosity of the $\alpha\si{ox} -$luminosity relation given by
Strateva et al. (2005, see also Yuan Brinkmann \& Siebert 1998) for
radio-quiet AGNs, similar to the result for a small sample of IMBH
AGN (Greene \& Ho  2006).

As results from the above HST image decomposition analysis (\S2.2),
the galaxy is composed of three components, a central point source
as the AGN, an inner S\'{e}rsic with $r_e = 0\arcsec.18$ (114\,pc),
and an outer S\'ersic with $r_e = 1\arcsec.58$ (1\,kpc), which
account for 6.0, 2.6, 91.4 per cent of the total F606W light,
respectively. We tend to identify the outer S\'ersic to be a
galactic disk, because of 1) its exponential-like radial profile
with the S\'ersic index $n=0.8$, close to 1 for exponential\footnote
{This somewhat shallower inner profile may be caused by dust
extinction in the disk, which is indicated by the Balmer decrements,
since the extinction aggravates inwards for an inclined disk ($i =
cos^{-1}(b/a)=66 \arcdeg$).}, 2) the disky isophote shape
($c=-0.1$), 3) the apparent extreme flattening (Hubble type E6.3
based on the axis ratio), that is very unlikely for an elliptical
(Binney \& de Vaucouleurs 1981), and particularly, 4) the existence
of the additional inner S\'ersic component.

With an effective radius of 0\arcsec.18, the inner S\'ersic is
merely resolved by the Wide Field Camera; however, fits without this
component result in large residuals and an over-predicted point
source, as displayed in Figure 2. Its size is too large for being a
nuclear star cluster. The fitted image has a boxy shape ($c=0.6$)
and a rather flat profile ($n=0.4$), suggesting a box-/peanut-shaped
bulge or an edge-on \emph{nuclear} bar---their presence in disk
galaxies are known for many years (eg. Burbidge \& Burbidge 1959;
Evans 1951; de Vaucouleurs 1975). We prefer the nuclear bar
speculation because there is a $\sim$17\arcdeg~mis-alignment between
its major axis and that of the disk, that is more natural for the
bar scenario; moreover, a bar can provide an efficient mechanism of
fueling gas to the AGN but a bulge cannot (e.g., Shlosman et al.
1989). In fact, a boxy bulge may evolve from a bar, or may simply be
(part of) a bar viewed edge-on (see Kormendy \& Kennicutt 2004 for a
review). Regardless of its nature, we predict the mass of the
central black hole based on the empirical $M_{BH} - L_{bul}$
correlations using the absolute magnitudes (Table 3, corrected for
the internal extinction $E_{B-V}=0.36$). Using the formalism from
Marconi \& Hunt (2003) and McLure \& Dunlop (2002), we find
$2.5\times 10^5$ and $2.6 \times 10^4$\msun, respectively, roughly
consistent with our estimate from the scaling relation considering
the large uncertainties. Furthermore, following the empirical
relation between $M_{BH}$ and bulge concentration (Graham et al.
2001, Graham et al. 2003), we obtain $M_{BH} = 1.6 \times
10^5$\msun, also consistent with the virial mass estimate. Either
bars or boxy bulges (as a kind of ``pseudobulges'' [Kormendy 1993])
form out of disk material (Kormendy \& Kennicutt 2004). So, probably
just as stated by Carollo (2004), ``the requirement for making a
supermassive central black hole is that the galaxy is capable of
reaching sufficiently high central baryonic densities.''

\acknowledgments This work is supported by Chinese NSF grants
NSF-10533050 and NSF-10573015. XBD is partially supported by a
postdoc grant from Wang Kuan-Cheng Foundation. We wish to thank the
anonymous referee for constructive comments and suggestions, and
thank Alister Graham for helpful comments. XBD thank Professor
Robert Kennicutt, Jr. for valuable suggestions and insightful
discussions during his visit to USTC this summer, and thank
Professor Zhenlong Zou for helpful comments and suggestions. Funding
for the Sloan Digital Sky Survey (SDSS)\footnote{ The SDSS Web site
is http://www.sdss.org/.} has been provided by the Alfred P. Sloan
Foundation, the Participating Institutions, the National Aeronautics
and Space Administration, the National Science Foundation, the U.S.
Department of Energy, the Japanese Monbukagakusho, and the Max
Planck Society. The SDSS is managed by the Astrophysical Research
Consortium (ARC) for the Participating Institutions.

\clearpage

\begin{deluxetable}{lccr}
\tablecaption{Fitted emission line parameters \label{tbl-1}}
\tablewidth{0pt} \tablehead{ \colhead{Line} &
\colhead{Centroid\tablenotemark{a}} &
\colhead{FWHM\tablenotemark{b}} & \colhead{Flux\tablenotemark{c}} }
\startdata

[O~II]$\lambda 3727$   &  3728.47  $\pm$  0.51                 &  281  $\pm$  102 &  28.0  $\pm$  8.5   \\
H$\beta$(narrow)\tablenotemark{d}     &  4862.14  $\pm$  0.24  &  137             &  17.0  $\pm$  2.9   \\
H$\beta$(broad)\tablenotemark{e}      &  4860.93  $\pm$  1.26  &  781             &  27.3  $\pm$  5.8   \\
$[O~III]\lambda 5007$  &  5008.25  $\pm$  0.08                 &  160  $\pm$  12  &  53.7  $\pm$  2.8   \\
$[O~I]\lambda 6301$    &  6302.83  $\pm$  0.86                 &  277  $\pm$  84  &  9.3   $\pm$  2.6   \\
H$\alpha$(narrow)     &  6564.53  $\pm$  0.06                  &  137  $\pm$   8  &  67.3  $\pm$  4.1   \\
H$\alpha$(broad)      &  6565.78  $\pm$  0.30                  &  781  $\pm$  42  &  115.2 $\pm$  5.3   \\
$[N~II]\lambda 6583$\tablenotemark{d}   &  6585.22             &  137             &  38.4  $\pm$  1.9   \\
$[S~II]\lambda 6716$   &  6718.43  $\pm$  0.18                 &  142  $\pm$  20  &  18.2  $\pm$  1.9   \\
$[S~II]\lambda 6731$\tablenotemark{f}   &  6732.81             &  142             &  14.6  $\pm$  1.9 %
\enddata
\tablecomments{(a): Vacuum rest-frame wavelengths, in unit of \AA.
(b): Corrected for the instrumental broadening; in
unit of km~s$^{-1}$. (c): In unit of $10^{-17}$\flux. (d): Adopting
the profile of narrow H$\alpha$. (e): Adopting the profile of broad
H$\alpha$. (f): Adopting the profile of [S~II]$\lambda6716$. }
\end{deluxetable}

\clearpage

\begin{deluxetable}{lccccccccc}
\tablewidth{0pt} \tablecaption {Two-Dimensional Image Fitting
Parameters} \tablehead{ Func. & m$_{606}$\tablenotemark{a} &
M$_{R}$\tablenotemark{a} & M$_{B}$\tablenotemark{a} & $r_{\rm{e}}$
& $n$ & $b/a$ & PA & $c$ & notes \\
(1)      & (2)    & (3)  & (4) & (5) & (6) & (7) & (8) & (9) & (10)
} \startdata
PSF      &  21.1 & $-14.6$ & $-14.6$ & --   & --   & --   & --   & --     & AGN      \\
S\'ersic &  21.9 & $-13.9$ & $-12.4$ & 0.18 & 0.4 & 0.28 & 82 & $+0.6$ & Bar?     \\
S\'ersic &  18.1 & $-17.8$\tablenotemark{b} & $-16.4$ & 1.58 & 0.8 & 0.37 & 65 & $-0.1$ & Disk%
\enddata
\tablecomments{ Col. (1): Components used in the fit. Col. (2): The
F606W integrated magnitudes on the Vega system. Col. (3): The
absolute Johnson R magnitudes. Col. (4): The absolute Johnson B
magnitudes. Col. (5): The effective radius of the S\'ersic law, in
arcsecond. Col. (6): The S\'ersic exponent. Col. (7): Axis ratio.
Col. (8): Position angle, in degree. Col. (9): Diskiness
(negative)/boxiness (positive) parameter, defined in equation (3) of
Peng et al. (2002). (a): All the magnitudes are corrected for the
Galactic extinction; see the text for the transformation from
mag$_{606}$ to Johnson B and V magnitudes. (b): In good agreement
with the SDSS $r-$band absolute magnitude derived from the fit with
an exponential model, $M_{r} = -17.8$.}

\end{deluxetable}

\clearpage

\begin{deluxetable}{lccc}
\tablewidth{0pt} \tablecaption {X-ray observations and data
reduction information} \tablehead{ &  XMM/EPIC-PN & ROSAT/HRI &
ROSAT/PSPC } \startdata

Obs. id              & 0147210301    & RH703861A01   & RP800517N00 \\
good exposure        & 5.7ks        & 54.9ks          & 12.1ks        \\
$R_{\mathrm{source~extraction}}$   & 30\arcsec   & 25\arcsec     & 45\arcsec   \\
net count rate\tablenotemark{*}    & $1.2\pm0.2$    & $0.12\pm0.02$  & $0.35\pm0.08$ \\
flux\tablenotemark{*}  & $3.4\pm0.6$ & $4.2\pm0.7$ &  $7.1\pm1.4$%

\enddata
\tablecomments{ (*): Measured in the 0.3--2.4keV range; count rate
in unit of $10^{-2}$\ucre, and flux (without Galactic absorption
correction), $10^{-14}$\flux. }

\end{deluxetable}


\clearpage

\begin{deluxetable}{llcccc}
\tablewidth{0pt} \tablecaption {Comparison with NGC 4395 and POX 52}
\tablehead{
Object & M$_{B}^{Host}$ & M$_{B}^{AGN}$ & $L_{H\alpha}^{br}$ & $L_{X,Soft}$ & $L_{X,Hard}$ \\
(1)      & (2)    & (3)  & (4) & (5) & (6) } \startdata
NGC 4395   & $-17.5$\tablenotemark{a}  &  $-10.8$\tablenotemark{a} & 1.2\tablenotemark{b} & 1.3\tablenotemark{c} & 1.5\tablenotemark{d}  \\
J1605+1748 & $-16.4$  &  $-14.6$                  & 27 & $1100$\tablenotemark{e}  & 7.0   \\
POX 52     & $-16.8$\tablenotemark{f}  &  $-16.7$\tablenotemark{f}                 & 120\tablenotemark{f} & -- & --%
\enddata
\tablecomments{Col. (4, 5): In unit of $10^{38}$ \lum. Col. (6):
Unabsorbed X-ray luminosity in the $2-10$keV range, in unit of
$10^{40}$ \lum. (a): From FH03. (b): From FS89. (c): Observed X-ray
luminosity in the $0.2-2$keV range from Moran et al. 1999. (d): From
Vaughan et al. 2005. (e): Mean observed X-ray luminosity in the
$0.3-2.4$keV. (f): Form B04. }
\end{deluxetable}

\clearpage

\begin{figure}
\plotone{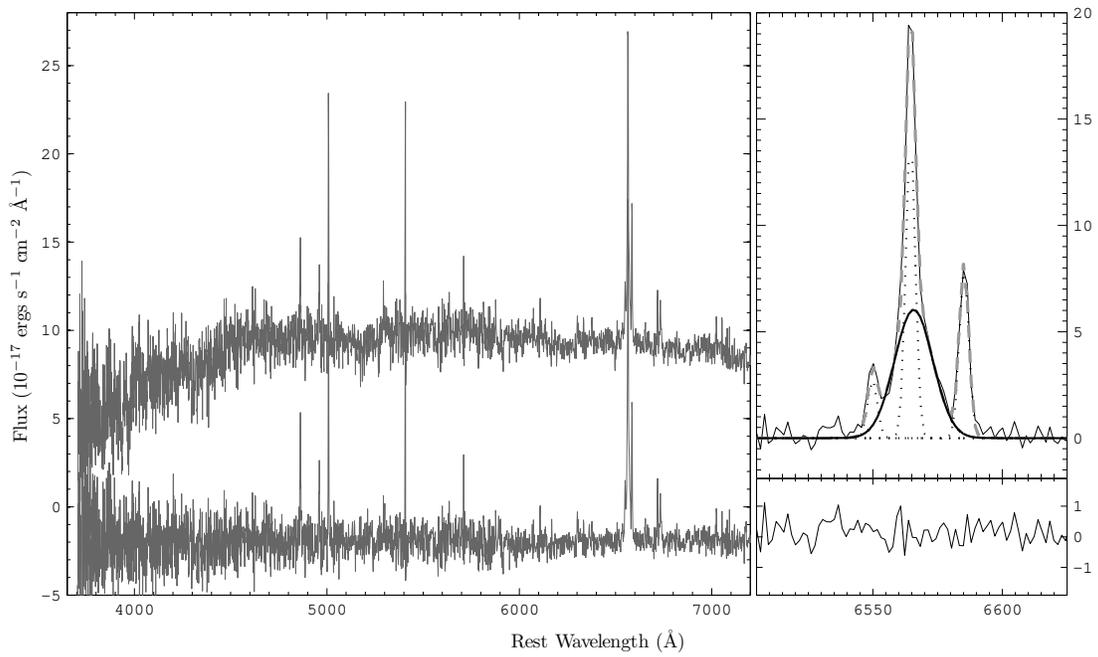} \caption{SDSS spectrum and the line fit.
Left: The SDSS spectrum (top) and the starlight/continuum subtracted
residual (bottom, offset downward by 2 units for clarity). Right:
The line decomposition in the H$\alpha +$[N II] region. The upper
panel shows the original data (thin solid line), the fitted narrow
lines (dotted lines), the fitted broad H$\alpha$ (thick solid line)
and the sum of the fit (gray dashed line). The lower panel, the
residual of the fit.
}
\end{figure}

\clearpage

\begin{figure}
\plotone{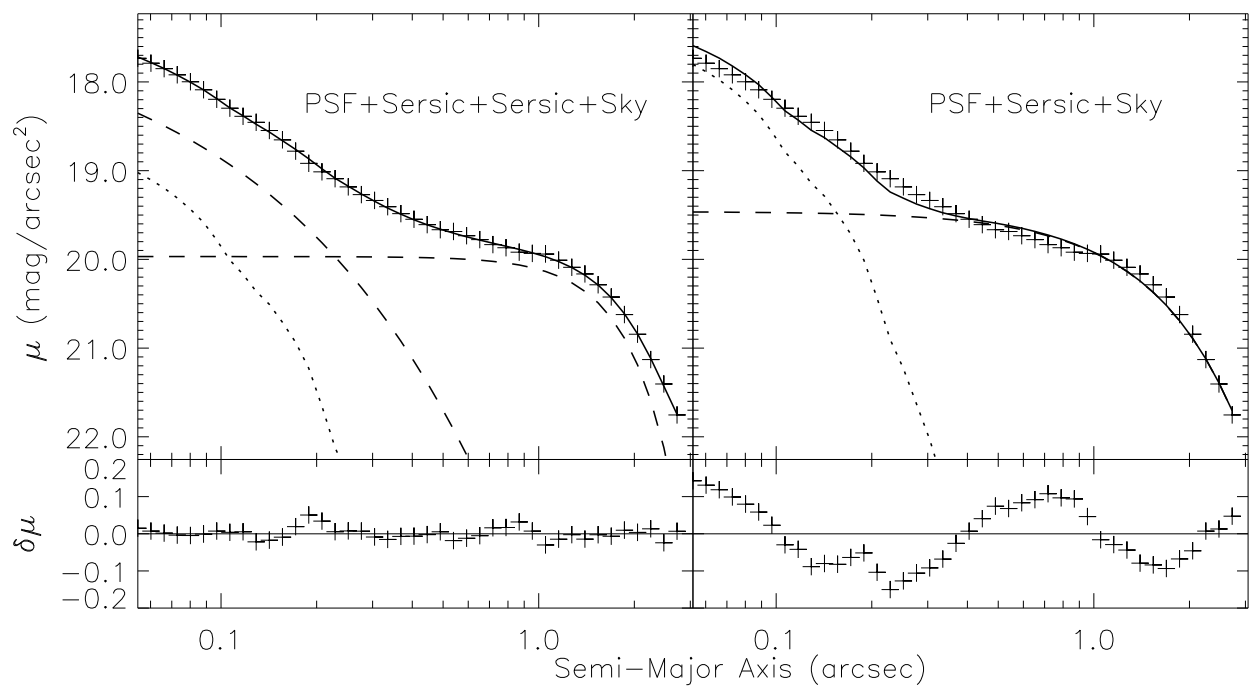} \caption{One dimensional fitting of the radial
profile, with PSF + two S\'ersic model (left panel) and PSF + a
S\'ersic model (right panel). Each panel shows the point-source
component (dotted), the host-galaxy component(s) (dashed), and the
sum (solid). Residuals are also displayed in the bottom. As
indicated in the right panel, fit with only one S\'ersic for the
host galaxy yields large residuals left and an over-predicted point
source component. The results of the 1-D fits are taken as initial
values of the fitting parameters of the subsequent 2-D
decomposition.}
\end{figure}

\clearpage

\begin{figure}
\plotone{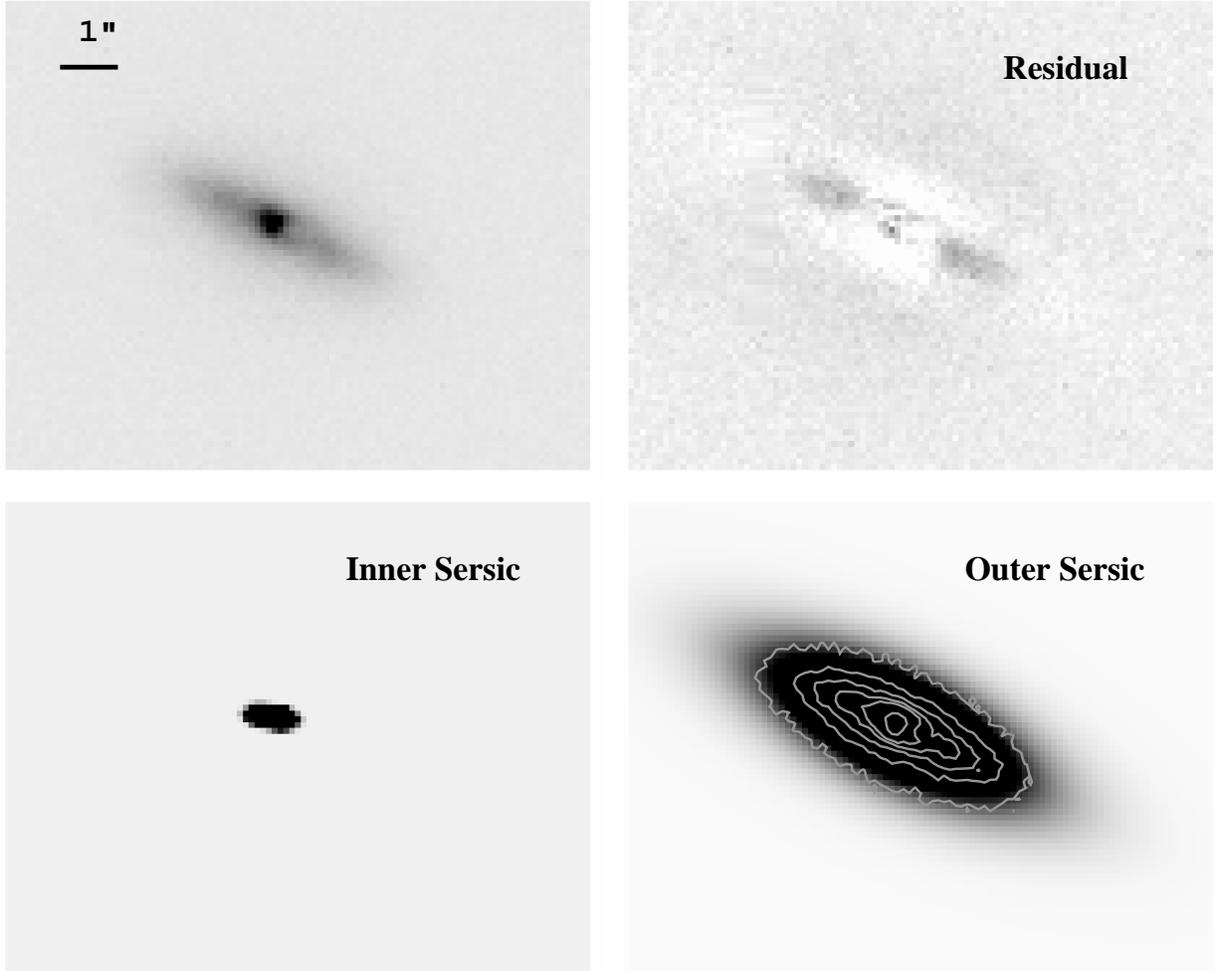} \caption{Two-dimensional decomposition of the
HST/WFPC2 F606W image. Top: the original data (left) and the
residual from the best-fitting PSF + two S\'ersic model (right).
The standard deviation of the distribution for the
fractional residuals $f_{residual}/f_{HST}\sim 0.1$ within the sky
region 10 times or more brighter than the sky level.
Bottom: the best-fit inner S\'ersic component (left) and outer
S\'ersic component with the contours of the original image overlaid
(right). Images have the same size. }
\end{figure}

\clearpage

\begin{figure}
\begin{flushleft} \textbf{\large Supplementary information:} \\
{\scriptsize [See more at
\textit{http://staff.ustc.edu.cn/\~{}xbdong/J1605+1748/}] }
\end{flushleft}

\plotone{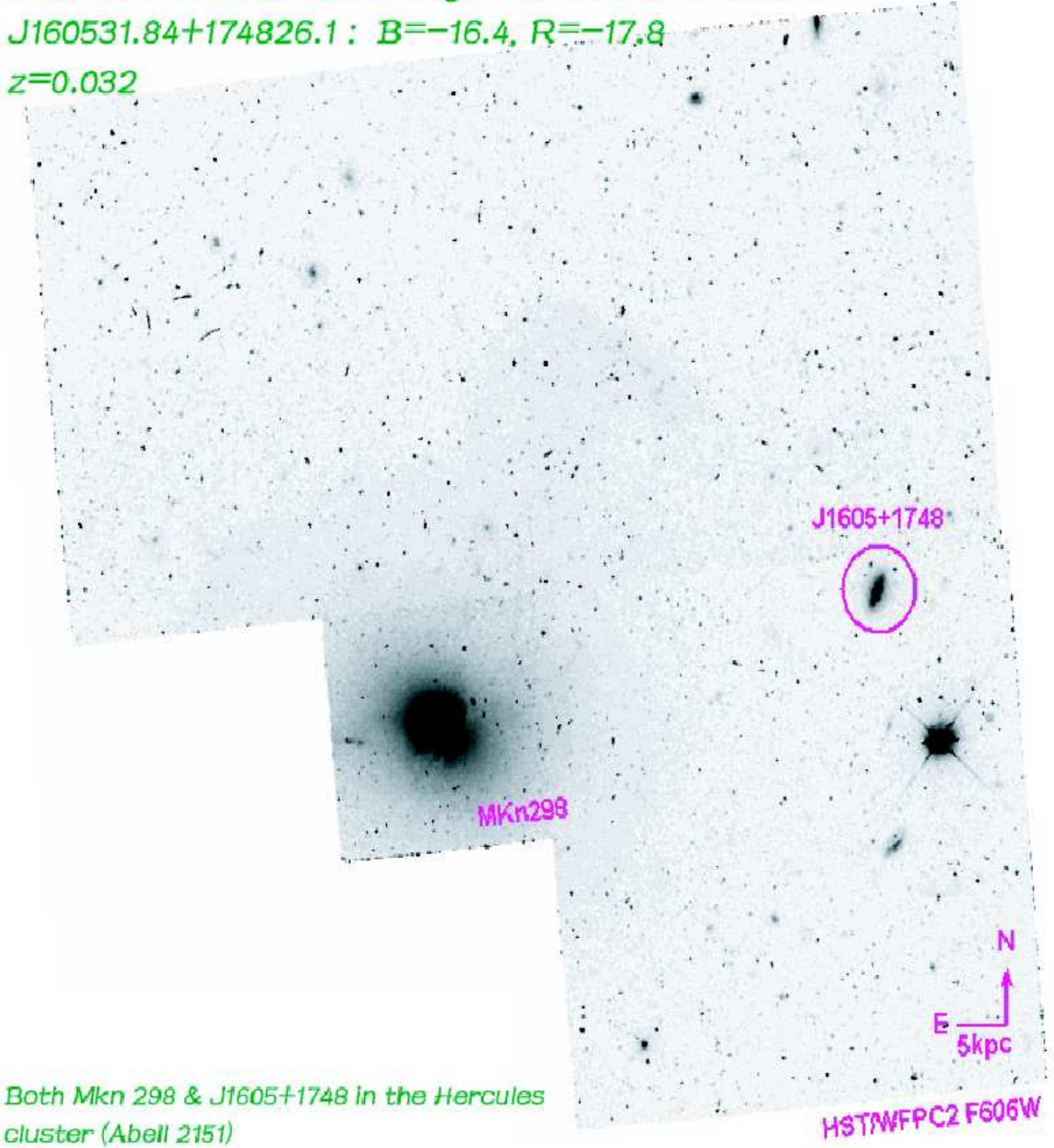} \caption{HST/WFPC2 F606W preview
image for J1605+1748 and its neighbor Mkn\,298. They are in the
Hercules cluster (Abell\,2151). Displayed with a logarithmic
stretch.}
\end{figure}

\end{document}